\documentclass[12pt,epsfig]{article}
\usepackage{graphicx}

\usepackage[latin1]{inputenc}
\usepackage{amsmath}
\usepackage{amsfonts}
\usepackage{amssymb}
\usepackage{graphicx}
\DeclareGraphicsExtensions{.pdf,.png,.jpg}

\usepackage{epstopdf}
\def\beq{\begin{equation}}
\def\eeq{\end{equation}}
\def\bea{\begin{eqnarray}}
\def\eea{\end{eqnarray}}

\begin{document}

\begin{center}
{\Large \bf Hartman effect from layered $PT$-symmetric system
  }

\vspace{1.3cm}

{\sf   Mohammad Hasan  \footnote{e-mail address: \ \ mhasan@isro.gov.in, \ \ mohammadhasan786@gmail.com}$^{,3}$,
 Bhabani Prasad Mandal \footnote{e-mail address:
\ \ bhabani.mandal@gmail.com, \ \ bhabani@bhu.ac.in  }}

\bigskip

{\em $^{1}$ Space Science Programme Office, Indian Space Research Organisation,
Bangalore-560094, INDIA \\
$^{2,3}$Department of Physics,
Banaras Hindu University,
Varanasi-221005, INDIA. \\ }

\bigskip
\bigskip

\noindent {\bf Abstract}

\end{center}
The time taken by a wave packet to cross through a finite layered $PT$-symmetric system is calculated by stationary phase method. We consider the $PT$- symmetric system of fix spatial length $L$ consisting of $N$ units of the potential system `$+iV$' and `$-iV$' of equal width `$b$' such that $L=2Nb$. In the limit of large `$b$', the tunneling time is found to be independent of $L$ and therefore the layered $PT$-symmetric system display the Hartman effect. The interesting limit of $N \rightarrow \infty$ such that $L$ remains finite is investigated analytically. In this limit the tunneling time matches with the time taken to cross an empty space of length $L$. The result of this limiting case $N \rightarrow \infty$ also shows the consistency of phase space method of calculating the tunneling time despite the existence of controversial Hartman effect. The reason of Hartman effect is unknown to present day however the other definitions of tunneling time that indicate a delay which depends upon the length of traversing region have  been effectively ruled out by recent attosecond measurements.

\medskip
\vspace{1in}
\newpage

\section{Introduction}        
The tunneling of a particle from  a classically forbidden region is one of the earliest studied problems of quantum mechanics  which started in the year $1928$ \cite{ nordheim1928, gurney1928}. Since then the quantum mechanical tunneling has long been studied by several authors \cite{condon, wigner_1955, david_bohm_1951}. However how much time does a particle take to tunnel through the barrier is an open question both theoretically and experimentally. In the year $1962$, Hartman applied the concept of stationary phase method to calculate the  time taken by a particle to cross a classically forbidden region. He considered the tunneling region imposed by metal-insulator-metal sandwich  and  shown that  tunneling time is independent of the barrier thickness for large barriers\cite{hartman_paper}. This paradox was later called as Hartman effect. This phenomena was also found by Fletcher in an independent study \cite{fletcher} in later years. The saturation of the tunneling time with barrier thickness is an obvious unexpected result as it contradict with the principles of special relativity. This has also created doubt on the methodology of stationary phase method of calculating the tunneling time and has prompted new definitions of tunneling time (see \cite{hg_winful}). However the  numerical calculation of the tunneling time by monitoring the time evolution of a particle wave packet have also indicated that the tunneling time agrees well with the one calculated by stationary phase method  \cite{aquino_1998}. 
\paragraph{}
Tunneling time have been studied for double and multi-barrier structures as well. The studies on double barrier structure have revealed the existence of generalized Hartman effect where the tunneling time is independent of the intervening gap of a double barrier system for large thickness \cite{generalized_hartman}. This also holds for multi-barrier case  \cite{esposito_multi_barrier}. See  \cite{questions_ghf1,questions_ghf2,questions_ghf3} for critical comments on generalized Hartman effect.   For the case when particle energy lies in the energy gap of super- lattice structure, the tunneling time through the super-lattice can be smaller than the free motion time \cite{pereyra_2000}. Ref \cite{pereyra_2000} also derives the close form expression of the tunneling time for super-lattice tunneling.  Owing to the generalization of standard quantum mechanics to non-Hermitian quantum mechanics, the particle tunneling have also been studied for complex barriers. Hartman effect doesn't exist in complex barrier tunneling \cite{complex_barrier_tunneling}. However the approach presented in \cite{complex_barrier_tunneling}  have been questioned in \cite{dutta} and a two channel formalism for incorporating inelasticity in barrier potential shows that Hartman effect exist for inelastic barriers with weak absorption \cite{dutta, our}. We have also calculated tunneling time in space fractional quantum mechanics (SFQM) and it is shown analytically that Hartman effect doesn't occur in SFQM \cite{tt_sfqm}.    
\paragraph{}
Experimental attempts have been made to test the finding of theoretical results about tunneling time. Many authors have contributed towards this. The earlier experiments have indicated the superluminal nature of the tunneling time \cite{sl_prl} and are found to be independent of the thickness of the tunneling region \cite {nimtz,ph,ragni,sattari,longhi1,olindo}.  The tunneling time was also studied with double barrier optical gratings \cite{longhi1} and double barrier photonic band gaps \cite{longhi2} and is found to be paradoxically short. Very recently, the attosecond measurement on one-electron tunneling  dynamics has indicated that the tunneling is an instantaneous phenomena \cite{angular_streaking_nature} within the experimental limitations of $1.8$ attoseconds. This  result effectively rule out the  other definitions of tunneling time which depends upon the traversal length (see \cite{angular_streaking_nature}).   
\paragraph{}
Around two decades ago, it was discovered that certain class of non-Hermitian Hamiltonian can support real energy eigen values provided the Hamiltonian is invariant under a combine parity and time-reversal symmetry. Since then a new dimension in quantum mechanics has emerged known as $PT$-symmetric quantum mechanics \cite {pt_book}. The non-Hermitian Hamiltonian display several new features which are originally absent in Hermitian Hamiltonians. The important features are exceptional points (EPs) \cite{ep1, ep2}, spectral singularity (SS) \cite{ss1, aop},coherent perfect absorption (CPA) \cite{cpa1}-\cite{cpa5}, critical coupling (CC) \cite{cc1}-\cite{cc4} and CPA-laser \cite{cpa_laser1}. Others notable features are invisibility  \cite{inv1, inv2, inv3} and reciprocity  \cite{resc}. Recently CPA and SS have also been studied in the context of non-Hermitian fractional quantum mechanics \cite{nh_sfqm}.     

In the present work, we investigate the existence of Hartman effect from a layered  $PT$-symmetric potential. We first take a `unit cell' $PT$-symmetric system of width $2b$ made by two complex rectangular barriers $iV$ and $-iV$ each of width $b$. We repeat this unit cell $N$ times without any intervening gap covering the total span $L=2Nb$. It is found that for $b \rightarrow \infty$ , the tunneling time is independent of $L$ which shows Hartman effect from this system. For a constant finite support $L$, each complex barriers become infinitely thin in the limit $N \rightarrow \infty$ (as $b=\frac{L}{2N}$). In this limit it is expected that the effect of adjacent barriers $iV$ and $-iV$ will cancel each other and the tunneling time should equate to free passage time by the particle traversing the length $L$ in vacuum. This is indeed the case and the proof is shown analytically. This also shows the consistency of phase delay method of computing the tunneling time.  We organize the paper as follows: In section \ref{intro_tt} we briefly mention the phase space methodology of tunneling time. Section \ref{tt_pt_system} provides the detail steps of calculating the tunneling time from our locally periodic $PT$-symmetric system. In this section we give the close form expression of the tunneling time. Sub-section \ref{hartman_case} and \ref{free_passage_case} discusses the limiting case to obtain  Hartman effect and free passage time. Finally the results are discussed in section \ref{results_discussions}.

\section{Stationary phase method of tunneling time}
\label{intro_tt}
In this section we briefly introduce the reader about the stationary phase  methodology of calculating the tunneling time \cite{dutta_roy_book}. In stationary phase  method, the tunneling time of a localized wave packet traversing a potential barrier is defined as the time difference between the incoming and the outgoing peak of the wave packet. Consider a normalized Gaussian wave packet $U_{k_{0}} (k)$ with mean momentum $\hbar k_{0}$.  The time evolution of the wave packet propagating to positive $x$ direction would be
\begin{equation}
\int U_{k_{0}} (k)e^{i(kx-\frac{Et}{\hbar})}dk
\label{localized_wave_packet}
\end{equation}
where $k=\sqrt{2mE}$. On traversing the potential barrier of width $L$, the transmitted wave packet would be
\begin{equation}
\int U_{k_{0}} (k) \vert A(k) \vert e^{i(kx-\frac{Et}{\hbar} +\theta(k))}dk
\label{emerged_wave_packet}
\end{equation}
where $A(k)=\vert A(k) \vert e^{i\theta (k)}$ is the transmission coefficient through the potential barrier $V(x)$ ($V(x)=V $ for $0 \leq x \leq L$ and zero elsewhere). According to stationary phase method the tunneling time $\tau$ is given by
\begin{equation}
\frac{d}{dk} \left( kL-\frac{E\tau}{\hbar} +\theta(k) \right)=0
\label{spm_condition}
\end{equation}
The tunneling time expression then becomes,
\begin{equation}
\tau= \hbar \frac{d \theta(E)}{dE} +\frac{L}{(\frac{\hbar k}{m})}
\label{phase_delay_time}
\end{equation}
For a square barrier potential $V(x)=V$ of width $L$, the tunneling time is
\begin{equation}
\tau= \hbar \frac{d}{dE} \tan^{-1} \left( \frac{k^{2}-q^{2}}{2kq} \tanh{qL}\right)
\end{equation}
Here $q= \sqrt{2m(V-E)}/\hbar$. It is seen that for $L \rightarrow 0$, we have $\tau \rightarrow 0$. This is expected, however for $L \rightarrow \infty$, $\tau=\frac{2m}{\hbar qk}$, This doesn't involve any dependency on $L$ i.e. tunneling time is independent of the width of the barrier $L$ for thick barrier. This is the famous Hartman effect.  In the system of unit  $2m=1$, $\hbar=1$, $c=1$
\begin{equation}
\lim_{L\rightarrow \infty} \tau =\frac{1}{qk}
\label{tt_qm}
\end{equation}

\section {Tunneling time for periodic $PT$-symmetric system}
\label{tt_pt_system}
As discussed in the previous section, the calculation of tunneling time by stationary phase method involves the calculation of phase of the transmission  coefficient and differentiation of the same with respect to energy of the particle to obtain phase delay time. The phase delay time when combined with the free passage time gives the net tunneling time of the particle. To achieve this we first calculate the transmission coefficient of the locally periodic $PT$-symmetric system by using transfer matrix method. For the sake of completeness, we also describe the transfer matrix method of calculating the transmission coefficient. The details of the methodology and calculations are illustrated in subsequent section.
\subsection{Transfer matrix} 
The Hamiltonian operator in one dimension for a non-relativistic particle is (in the unit $\hbar=1$ and $2m=1$)
\begin{equation}
H=-\frac{d^{2}}{d x^{2}}+ V(x)
\label{hamiltonian_operator}
\end{equation}
where $V(x) \in C$ . $V(x) \rightarrow 0$ as $x \rightarrow \pm \infty$. If $\int U (x) dx$, where $U(x)=(1+\vert x \vert) V(x)$ is finite over all $x$, then the Hamiltonian given above admits a scattering solution with the following asymptotic values
\begin{eqnarray}
\psi (k,x \rightarrow +\infty)= A_{+}(k) e^{ikx}+B_{+}(k) e^{-ikx} \\
\psi (k,x \rightarrow -\infty)= A_{-}(k) e^{ikx}+B_{-}(k) e^{-ikx}
\end{eqnarray}    
The coefficients $A_{\pm}, B_{\pm}$ are connected through a $2 \times  2$ matrix $M$, called as transfer matrix as given below,
\beq
\begin{pmatrix}   A_{+}(k) \\ B_{+}(k)     \end{pmatrix}= M(k) \begin{pmatrix}   A_{-}(k) \\ B_{-}(k)    \end{pmatrix} 
\eeq
where,
\beq
 M(k)= \begin{pmatrix}   M_{11}(k) & M_{12}(k) \\ M_{21}(k) & M_{22}(k)   \end{pmatrix}  
\eeq
With the knowledge of the transfer matrix $M(k)$, the transmission coefficient $t(k)$ is easily obtained as the inverse of the lower diagonal element (for wave incidence from left of the potential) i.e.
\begin{equation}
t(k)=\frac{1}{M_{22}(k)}
\label{tl_general}
\end{equation} 
The transfer matrix shows composition property. If the transfer matrix for two non-overlapping finite scattering regions $V_{1}$ and $V_{2}$ , where $V_{1}$ is to the left of $V_{2}$, are $M_{1}$ and $M_{2}$ respectively, then the net transfer matrix $M_{net}$ of the whole system ($V_{1}$ and $V_{2}$) is
\beq
M_{net}=M_{2}. M_{1}
\eeq 
The composition result can be generalized for arbitrary numbers of non-overlapping finite scattering regions. Knowing the transfer matrix, one easily compute the scattering coefficients by inversing the lower diagonal element of the net transfer matrix. 
\subsection{Transmission coefficient of locally periodic $PT$-symmetric system and tunneling time}
\begin{figure}
\begin{center}
\includegraphics[scale=0.5]{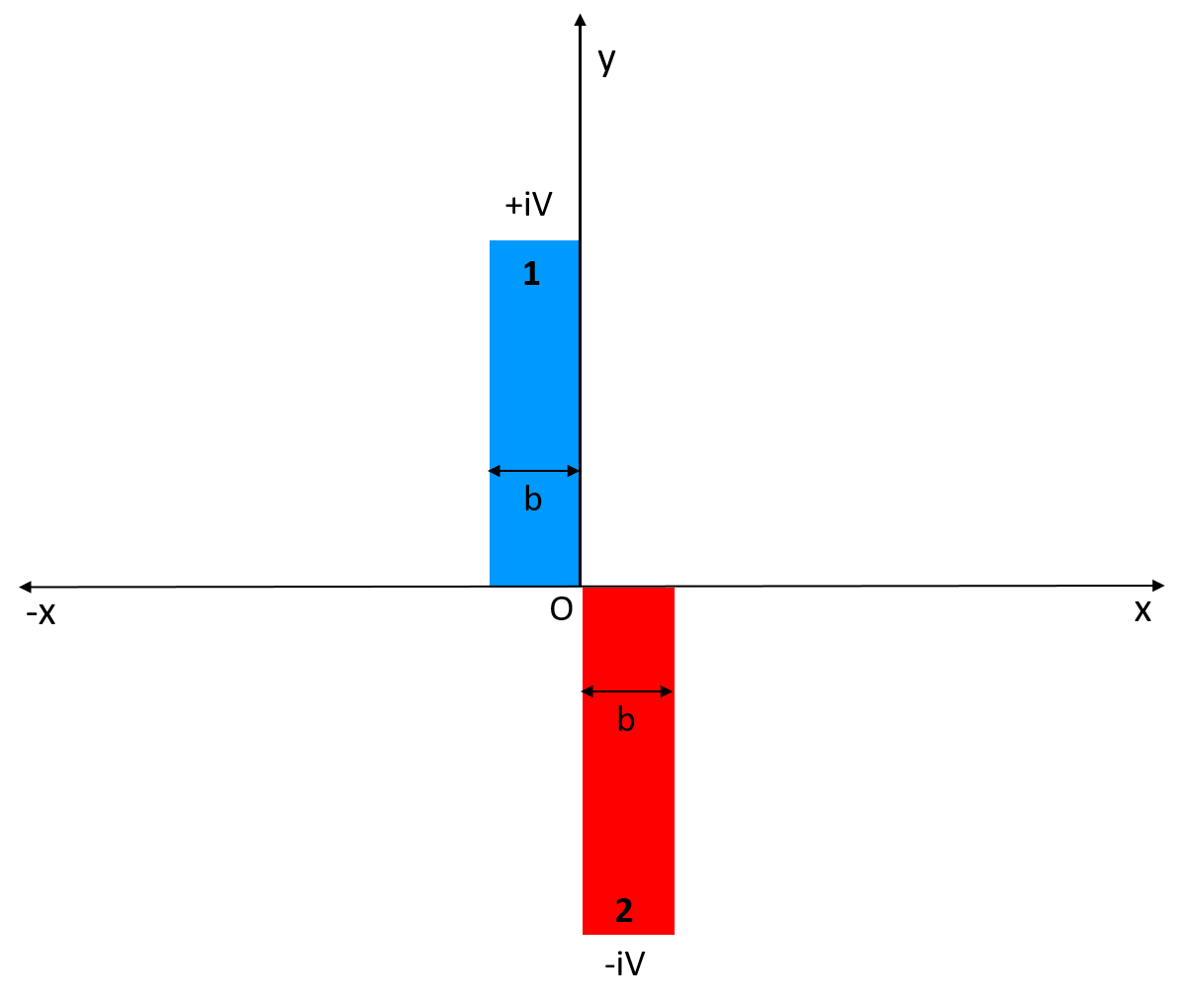}  
\caption{\it A $PT$-symmetric `unit cell' consisting a pair of complex conjugate barrier. $y$-axis represent the imaginary  height of the potential.}  
\label{pt_barrier}
\end{center}
\end{figure}

Fig \ref{pt_barrier} represent our $PT$-symmetric `unit cell' barrier configuration. The transfer matrices for the two barriers labeled as `$1$' and `$2$' are
\beq
 M_{1,2}(k)= \frac{1}{2 }\begin{pmatrix}   e^{-ikb} P_{+}^{1,2} & e^{-ikb(1+2j)} S^{1,2} \\ -e^{ikb(1+2j)} S^{1,2} & e^{ikb} P_{-}^{1,2}   \end{pmatrix}  
\eeq 
In the above $j=0$ for barrier-$1$ and $j=1$ for barrier-$2$ as labeled in Fig \ref{pt_barrier} and,
\beq
P_{\pm}^{1,2}=2 \cos{k_{1,2} b} \pm i \left(\mu_{1,2} +\frac{1}{\mu_{1,2}} \right) \sin{k_{1,2} b}
\label{p_expression}
\eeq
\beq
S^{1,2}= i \left(\mu_{1,2} -\frac{1}{\mu_{1,2}} \right) \sin{k_{1,2} b}
\label{s_expression}
\eeq
\beq
\mu_{1,2}=\frac{k_{1,2}}{k}, \ \  k_{1,2}=\sqrt{E-V_{1,2}}
\label{u12}
\eeq
\begin{figure}
\begin{center}
\includegraphics[scale=0.5]{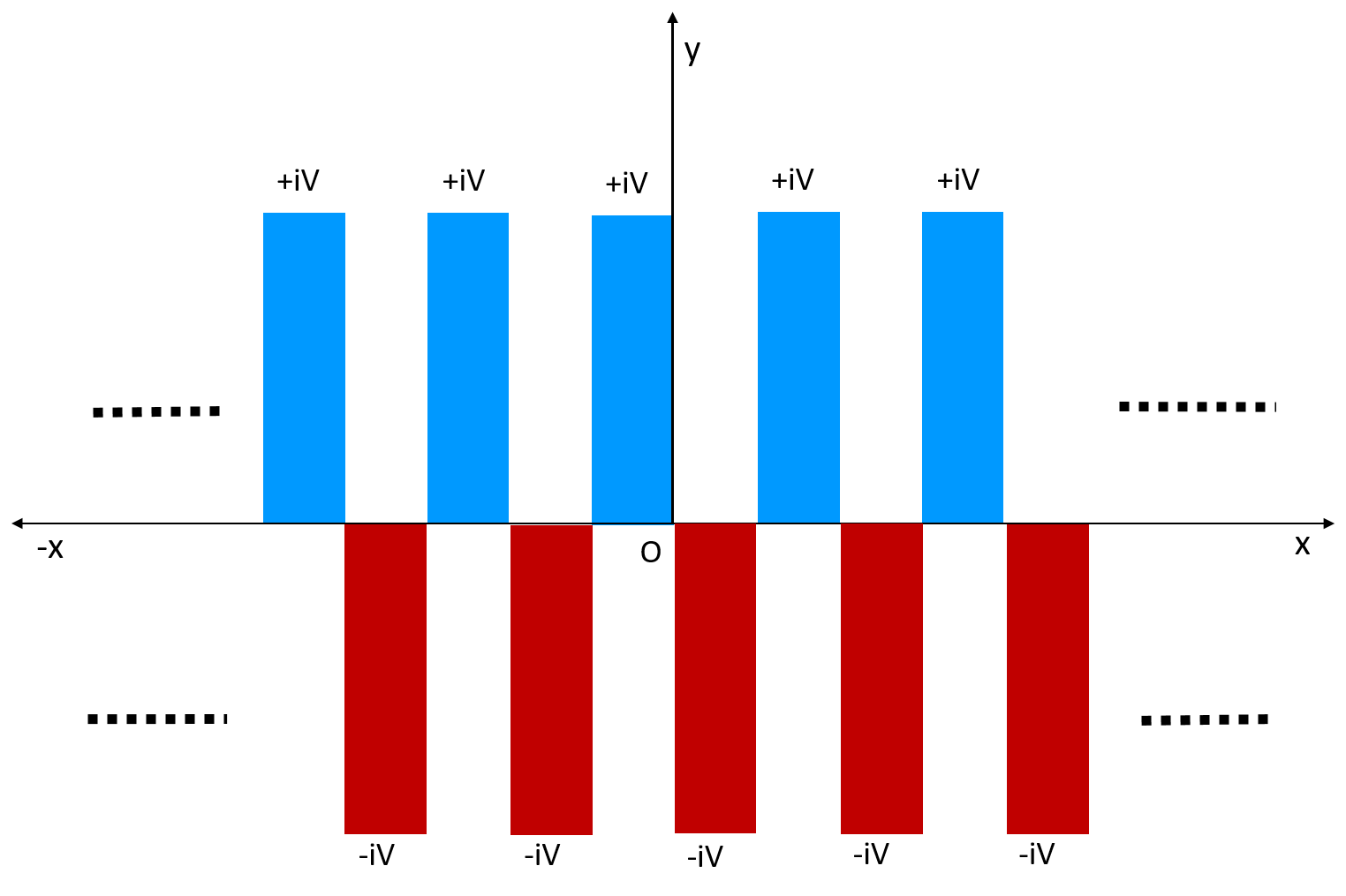}
\caption{\it A periodic $PT$-symmetric potential made by the periodic repetition of the `unit cell' potential shown in Fig \ref{pt_barrier}. $y$-axis represent the imaginary height of the potential.}  
\label{periodic_pt_barrier}
\end{center}
\end{figure}  
For the present problem $V_{1}=iV$ and $V_{2}=-iV$. By using the composition properties of the transfer matrix, the net transfer matrix of our $PT$-symmetric `unit cell' can be easily found as 
\beq
 M(k)= \frac{1}{4 }\begin{pmatrix}   e^{-2ikb} (P_{+}^{1} P_{+}^{2}-S^{1}S^{2}) & e^{-2ikb} (P_{+}^{2} S^{1}+P_{-}^{1} S^{2}) \\ -e^{2ikb} (P_{-}^{2} S^{1}+P_{+}^{1} S^{2}) & e^{2ikb} (P_{-}^{1} P_{-}^{2}-S^{1}S^{2})   \end{pmatrix} 
\label{pt_transfer} 
\eeq 
If the transfer matrix of the `unit cell' potential is known, one can obtain the transfer matrix of the corresponding locally periodic \cite{griffith_periodic} as well as locally super periodic potential \cite{super_periodic} and therefore the transmission coefficient from such a system is easily obtained. In Fig \ref{periodic_pt_barrier} we show the locally periodic $PT$-symmetric potential obtained by periodic repetition of our $PT$-symmetric `unit cell'. For the system to be $PT$-symmetric, there required to be total $n$ barriers ( combined $iV$ and $-iV$ barriers and not the `unit cell') on either side of the origin. Thus the total barriers are  $n=2N$, $N \in I^{+}$ and the net `unit cells' are $N$. On doing the algebraic simplification of the transfer matrix \ref{pt_transfer} for the potential $V_{1}=iV$ , $V_{2}=-iV$ by using eqs. \ref{p_expression}, \ref{s_expression}, \ref{u12} and following the method outlined in \cite{griffith_periodic} we obtain the close form expression of the transmission coefficient. Below we write the final expression of transmission coefficient from our locally periodic $PT$-symmetric system as 
\beq
t=\frac{e^{-ikL}}{G(k)}
\label{t_simple}
\eeq
where $G(k)$ is given as
\beq
G(k)=(\xi-i\chi ) U_{N-1}(\xi)-U_{N-2}(\xi)
\eeq
$U_{N}(\xi)$ is Chebyshev polynomial of second kind and $L=2Nb$ is the net spatial extent of the periodic potential. $\{\xi, \chi \} \in R$. The expressions for $\xi$ and $\chi$ are given below 
\beq
\xi= \frac{1}{2}(\cos{2\alpha}+\cosh{2\beta}) -\cos{2\phi} (\cosh^{2}{\beta} \sin^{2}{\alpha}+ \cos^{2}{\alpha} \sinh^{2}{\beta})
\label{a_eq}
\eeq
\beq
\chi= \frac{1}{2} (U_{+} \cos{\phi} \sin{2\alpha} + U_{-} \sin{\phi} \sinh{2\beta})
\label{b_eq}
\eeq
In the above equations, $U_{\pm}= \frac{k}{\rho} \pm \frac{\rho}{k}$, $\alpha=b \rho \cos{\phi}$, $\beta=b \rho \sin{\phi}$. $\rho$ and $\phi$ are the modulus and phase of $k_{2}=\sqrt{k^{2}+iV}= \rho e^{i \phi}$ respectively such that $\rho=(k^{4}+V^{2})^{\frac{1}{4}}$ and $\phi= \frac{1}{2}\tan^{-1} \left( \frac{V}{k^{2}} \right )$. It can be noted that $k_{1}= \rho e^{-i \phi}$. The detail exercise of the above is left to the interested reader. To find the tunneling time, we now separate $G(k)$ in real and imaginary parts and calculate the phase $\theta$ of the transmission coefficient $t$ (eq. \ref{t_simple}). The expression of $\theta$ is calculated to be
\beq
\theta= \tan^{-1} ( q \chi   ) -kL
\label{theta_eq}
\eeq   
Here,
\beq
q=\frac{U_{N-1} (\xi)}{T_{N} (\xi)}
\label{q_eq}
\eeq
Using $\theta$ in eq. \ref{phase_delay_time} the final expressions for the tunneling time from our $PT$-symmetric system is found to be
\beq
\tau= \frac{1}{2k(1+ q^{2}\chi^{2})} \left [q \chi' +  \chi \left ( \frac{N \xi'}{\xi^{2}-1} - N \xi' q^{2}- \frac{q \xi \xi'}{\xi^{2}-1}  \right ) \right ]
\label{time_eq}
\eeq
In the above $\prime$ denote the derivative with respect to $k$. The expression for $\xi'$ and $\chi'$ are
\beq
\xi'=2 \beta' \sin^{2}{\phi} \sinh{2\beta} -2 \alpha' \cos^{2}{\phi} \sin{2\alpha} + \phi' \sin{2\phi} (\cosh{2\beta}-\cos{2\alpha})
\label{xi_prime}
\eeq
\begin{multline}
\chi'=U_{+} (\alpha' \cos{2\alpha} \cos{\phi}- \frac{1}{2}\phi' \sin{\phi} \sin{2\alpha})+U_{-} (\beta' \cosh{2\beta}\sin{\phi}+\frac{1}{2}\phi' \cos{\phi}\sinh{2\beta}) \\ + \frac{1}{2}U_{+}' \cos{\phi} \sin{2\alpha}+ \frac{1}{2} U_{-}' \sin{\phi} \sinh{2\beta} 
\label{chi_prime}
\end{multline}
Other quantities with prime are defined below,
\beq
U_{\pm}'= \frac{V^{2}}{\rho^{5}} \left (1 \mp \frac{\rho^{2}}{k^{2}} \right ) , \ \ \rho'=\left ( \frac{k}{\rho}\right  )^{3}, \ \ \phi'=-\frac{kV}{\rho^{4}}
\nonumber
\eeq
\beq
\alpha'=\frac{b k}{\rho^{3}}(V \sin{\phi}+k^{2}\cos{\phi}), \ \ \beta'=\frac{b k}{\rho^{3}}(-V \cos{\phi}+k^{2}\sin{\phi}) 
\nonumber
\eeq
\subsection{Special case 1: Existence of Hartman effect}
\label{hartman_case}
\begin{figure}
\begin{center}
\includegraphics[scale=0.45]{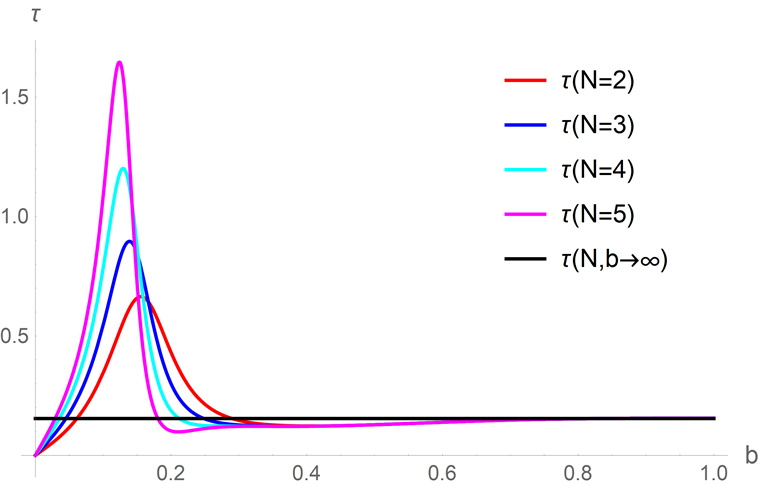}
\caption{\it Variation of tunneling time $\tau$ with width `$b$' of the `unit cell' barrier for different values of $N$. The black line is drawn at the theoretical value of $\tau$  in the limit $b \rightarrow \infty$ obtained in  eq. \ref{time3}. It is seen that all curve of different $N$ approaches to this value with increasing $b$. In the figure $V=20$ and $E=1$.} 
\label{hartman_fig}
\end{center}
\end{figure}

To study the existence of Hartman effect we take the limiting case $b \rightarrow \infty$ of the tunneling time $\tau$ and evaluate the dependency on $b$ in this limiting case. To arrive at the final result we evaluate the limiting case of the following quantities
\beq
\lim _{b \rightarrow \infty} \xi \sim  f_{1} e^{2 \beta}
\label{xi_infinite}
\eeq    
\beq
\lim _{b \rightarrow \infty} \xi' \sim (f_{2}+b f_{4}) e^{2 \beta} +b f_{3}
\label{xi_prime_infinite}
\eeq    
\beq
\lim _{b \rightarrow \infty} q \sim \frac{1}{\xi (b \rightarrow \infty)}
\label{q_infinite}
\eeq    
\beq
\lim _{b \rightarrow \infty} \chi \sim \frac{1}{4} U_{-} e^{2 \beta} \sin{\phi} 
\label{chi_infinite}
\eeq    
\beq
\lim _{b \rightarrow \infty} \chi' \sim b g_{1} +(b g_{2}+g_{3}) e^{2 \beta}
\label{chi_prime_infinite}
\eeq    
\beq
\lim _{b  \rightarrow \infty} \frac{\chi}{\xi} = \gamma
\label{chi_xi_ratio_infinite}
\eeq    
In eqs \ref{xi_infinite}-\ref{chi_xi_ratio_infinite}, we have used `$\sim$' when the right hand side (RHS) depends upon `$b$'. In case when `$b$' doesn't appear in RHS, we have used `$=$' sign.  Various quantities appearing in the eqs. \ref{xi_infinite}- \ref{chi_xi_ratio_infinite}  are defined below.
\beq
f_{1}= \frac{1}{2} \sin^{2}{\phi}
\label{f1_eq}
\eeq
\beq
f_{2}= \frac{1}{2} \phi' \sin{2\phi}
\label{f1_eq}
\eeq
\beq
f_{3}= -\frac{2k}{\rho^{3}} \sin{2\alpha}  \cos{\phi}( k^{2} \cos^{2}{\phi}+ \frac{1}{2} V \sin{2\phi}) 
\label{f3_eq}
\eeq
\beq
f_{4}= \frac{k \sin{\phi}}{\rho^{3}} ( k^{2} \sin^{2}{\phi}- \frac{1}{2} V \sin{2\phi}) 
\label{f4_eq}
\eeq
\beq
g_{1}= \frac{kU_{+} \cos{2\alpha}}{\rho^{3}} (k^{2} \cos^{2}{\phi} +\frac{1}{2} V \sin{2\phi})  
\label{g1_eq}
\eeq
\beq
g_{2}=  \frac{kU_{-}}{2\rho^{3}}  (k^{2} \sin^{2}{\phi} -\frac{1}{2} V \sin{2\phi})  
\label{g2_eq}
\eeq
\beq
g_{3}= \frac{1}{4}(\phi' U_{-} \cos{\phi} + U_{-}' \sin{\phi})  
\label{g3_eq}
\eeq
\beq
\gamma= \frac{1}{2} U_{-}\csc{\phi} 
\label{gamma_eq}
\eeq
The quantities defined from eqs \ref{f1_eq}- \ref{gamma_eq} are independent of `$b$'. Due to $\lim_{b \rightarrow \infty} q \sim \frac{1}{\xi}$ and $\xi^{2} >> 1$ when $b \rightarrow \infty$, the first and second term of the parenthesis in the eq. \ref{time_eq} cancels and we are left with
\beq
\lim _{b \rightarrow \infty} \tau= \frac{1}{2k} \lim_{b \rightarrow \infty} \left [ \left (\frac{1}{1+ \gamma^{2}} \right ) \left ( \frac{\chi'}{\xi}- \frac{\chi \xi'}{\xi^{2}} \right )\right ]
\label{time1}
\eeq
From eq. \ref{time1} , it is evident that the tunneling time is independent of the repetitions $N$ of the unit cell in the limit $b \rightarrow \infty$ i.e. independent of the net spatial extent. This is Hartman effect. We further illustrate that the right hand side of eq. \ref{time1} is independent of the width `$b$' as well. Eq. \ref{time1} can be written as
\beq
\lim _{b \rightarrow \infty} \tau= \frac{1}{2k (1+ \gamma)} \left [ \frac{g_{3}-\gamma f_{2}}{f_{1}} + \frac{b}{f_{1}} (g_{2}-\gamma f_{4}) \right ] 
\label{time2}
\eeq
where we have used eqs \ref{xi_infinite}, \ref{xi_prime_infinite}, \ref{xi_prime_infinite} and \ref{chi_xi_ratio_infinite} in arriving at eq. \ref{time2}. From the expressions of $g_{2}$, $\gamma$ and $f_{4}$ it can be easily shown that
\beq
g_{2}-\gamma f_{4}=0
\eeq   
Thus,
\beq
\lim _{b \rightarrow \infty} \tau= \frac{1}{2k (1+ \gamma^{2})} \left [ \frac{g_{3}-\gamma f_{2}}{f_{1}}  \right ] 
\label{time3}
\eeq
Right hand side of the above equation is independent of the net span $L=2Nb$ of the potential as $\gamma$, $f_{1}$, $f_{2}$ and $g_{3}$ are independent of $b$. This proves the Hartman effect for our $PT$-symmetric system. This is also shown graphically for different $N$ values in Fig \ref{hartman_fig}.

\subsection{Special case: Free propagation}
\label{free_passage_case}
\begin{figure}
\begin{center}
\includegraphics[scale=0.45]{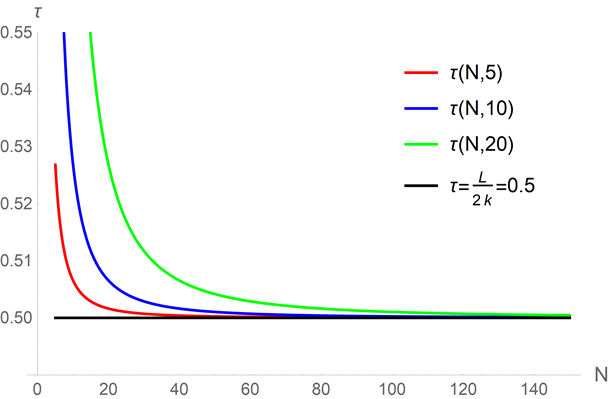} i \ \ \includegraphics[scale=0.45]{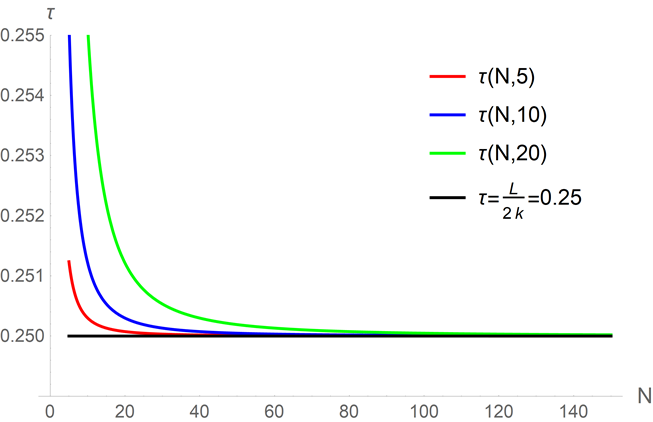} ii 
\caption{ \it Variation of tunneling time $\tau$ with the repetitions `$N$' of the $PT$-symmetric barrier for fixed spatial extent $L$. The horizontal black curve is drawn at the value $\tau= \frac{L}{2k}$. The curves correspond to different value of $V$ shown in the figure. The Fig-i is for $E=1$ and Fig-ii is for $E=4$. It is observed that with increasing $N$, all curves approach to the free propagation time $\frac{L}{2k}$ i.e. the time taken by a particle of wave vector $k$ to travel an empty region $L$.} 
\label{frp_fig}
\end{center}
\end{figure}
The free propagation limit is expected  when the width of each barriers are infinitely thin and number of the barriers are infinitely large such that the net span $L=2Nb$ of the $PT$-symmetric potential is fixed. In this limiting case the gain `$+iV$' and loss `$-iV$' part are expected to cancel each other and the net medium of length $L$ is free from any net gain or loss. In this case the tunneling time of a particle will approach to the free propagation time $L/2k$. To illustrate this , we take $b=\frac{L}{2N}$ where $L$ is fixed and study the limiting case of $N \rightarrow \infty$ in eq. \ref{time_eq}. Taking the limiting case of $N \rightarrow \infty$, it can be shown that
\beq
\lim_{N \rightarrow \infty} \tau=\frac{L}{4k \rho^{3}} \left [ (k^{3}U_{+}+\rho^{4} U_{+}')\cos^{2}{\phi}+(k^{3}U_{-}+\rho^{4}U_{-}')\sin^{2}{\phi} +\frac{1}{2}(U_{+}-U_{-}) (kV \sin{2\phi} -\phi' \rho^{4})  \right ]
\label{t_limit1}
\eeq
Using the expressions of $U_{+}, U_{-}$ and $U_{+}', U_{-}'$,  derived earlier this further simplifies to 
\beq
\lim_{N \rightarrow \infty} \tau=\frac{L}{4k \rho^{3}} \left [ \rho^{3}+ \left ( \frac{k^{4}-V^{2}}{k^{2}}\right ) \rho \cos{2 \phi}+2 V\rho \sin{2\phi} \right ]
\label{t_limit2}
\eeq 
Using $\sin{2\phi}=\frac{V}{\rho^{2}}, \ \cos{2\phi}=\frac{k^{2}}{\rho^{2}}$ in the above equation, the term in the square  parenthesis simplifies to $2 \rho^{3}$ and therefore,
\beq
\lim_{N \rightarrow \infty} \tau=\frac{L}{2k } \ \ \ , \mbox{for fixed $L$}
\label{t_limit3}
\eeq 
 This is precisely the free propagation time of a particle with wave vector $k$ traversing the free length $L$ (without medium). We have demonstrated this graphically in Fig \ref{frp_fig}.
\section{Results and Discussions}
\label{results_discussions}
In the present work, we have studied the tunneling time from a layered $PT$-symmetric system by stationary phase method.  We consider the $PT$- symmetric system of fix spatial length $L$ consisting of $N$ units of the potential system `$+iV$' and `$-iV$' of equal width `$b$' such that $L=2Nb$.  We have derived closed form expression of tunneling time for such a system for arbitrary finite repetitions $N$ . It is shown analytically and also demonstrated graphically that in the limit of large $b$, the tunneling time is independent of $b$ as well as the repetitions $N$. Thus the Hartman effect exist for layered $PT$-symmetric system. 
\paragraph{}
We have chosen our $PT$-symmetric system in such a way that for a fixed spatial extent `$L$' of the system, the `unit cell' $PT$-symmetric system becomes infinitely thin when the number of repetitions `$N$' is infinitely large. In this particular limit the whole system is expected to resemble an empty space of length $L$ and the tunneling time would approach the free propagation time $\frac{L}{2k}$ (chosen unit $2m=1$, $\hbar=1$, $c=1$) where $k$ is the wave vector of the particle. This is indeed the case and is proven analytically. We call this limiting case as free propagation limit and is demonstrated graphically as well. The analytical proof of free propagation limit also indicates the consistency of stationary phase method of calculating the tunneling time. To the best of our knowledge ,  tunneling time obtained by stationary phase method are the only known time  which shows insensitivity of barrier thickness for large barriers and thus tunneling can be instantaneous. The recent attosecond measurement have indicated  the instantaneous nature of tunneling of single electron system and thus have ruled out the other used definitions of tunneling time which are sensitive to the thickness of the tunneling region. It is an extraordinary fact to see that the mathematically complicate expressions of the derived tunneling time by stationary phase method for finite layered $PT$-symmetric system correctly reduces to free propagation time for length $L$ in the limit $N \rightarrow \infty$ for fix $L$. 

{\it \bf{Acknowledgements}}: \\
MH acknowledges supports from Director-SSPO for the encouragement of research activities. BPM acknowledges the support from CAS, Department of Physics, BHU.

\end{document}